\input{aipcheck}

\documentclass[
    ,final            
  ]
  {aipproc}

\layoutstyle{6x9}

\begin{document}

\title{Nucleon Resonance Electrocouplings from the CLAS Meson Electroproduction Data}


\author{I.~G.~Aznauryan}{
  address={Yerevan Physics Institute, 375036 Yerevan, Armenia}
}

\author{V.~D.~Burkert}{
  address={Thomas Jefferson National Accelerator Facility, 23606 Newport News, VA, USA}
}

\author{V.~I.~Mokeev}{
  address={Thomas Jefferson National Accelerator Facility, 23606 Newport News, VA, USA},
  email={mokeev@jlab.org},
  altaddress={Skobeltsyn Nuclear Physics Institute at Moscow State University, 1198899 Moscow, Russia}
}

\begin{abstract}
 Transition helicity amplitudes $\gamma_{v}NN^*$
(electrocouplings) were determined for prominent excited proton states with masses below 1.8 GeV in independent
analyses of major meson electroproduction channels: $\pi^{+}n$, $\pi^0p$ and $\pi^+\pi^-p$. 
Consistent results on
resonance electrocouplings obtained from analyses of these exclusive reactions with different non-resonant contributions 
demonstrate reliable extraction of these fundamental quantities for states that have significant
decays for either $N\pi$ or $N\pi\pi$ channels. 
Preliminary results on electrocouplings of 
$N^*$ states with masses above 1.6 GeV have become available from the CLAS data on $\pi^+\pi^-p$ electroproduction off protons for
the first time. 
Comparison  with quark models and coupled-channel approaches strongly suggest that $N^*$ structure is
determined by contributions from an internal core of three constituent quarks and an external meson-baryon cloud
at the distances covered in these measurements with the CLAS detector. 
\end{abstract}

\maketitle


\section{Introduction}

Studies of nucleon resonance structure in exclusive meson electroproduction off protons represent an important
direction in the $N^*$ program with the CLAS detector \cite{Bu11}, with the primary objective of determining
electrocouplings, of most excited proton states at photon virtualities $Q^2$ up to 5.0 GeV$^2$. 
This information allows
us to pin down active degrees of freedom in resonance structure at various distances, and eventually to access strong interaction mechanisms
that are responsible for $N^*$ formation from quarks and gluons \cite{Bu11,CL12,CRo11}. In this paper 
we report the 
results on the studies of  $\gamma_{v}NN^*$ electrocouplings of prominent excited proton states 
in the mass range up to 1.8
GeV from independent analyses of major meson electroproduction channels: 
$\pi^{+}n$, $\pi^0p$ and $\pi^+\pi^-p$. These channels are sensitive to resonance contributions, 
and they have different non-resonant mechanisms.  
Successful description of a large body of observables measured in  $\pi^{+}n$, $\pi^0p$ and 
$\pi^+\pi^-p$ 
electroproduction reactions, achieved with consistent values of $\gamma_{v}NN^*$ electrocouplings,
demonstrates the reliable
extraction of these fundamental quantities. Analysis of the results on the $\gamma_{v}NN^*$ 
electrocouplings 
open access to active degrees of freedom in $N^*$ structure at distances that correspond to the
 confinement regime at large values of the running quark-gluon coupling.

\section{The CLAS data on pion electroproduction off protons and analysis tools \label{tools}}

The CLAS data considerably extended information available on $\pi^{+}n$, $\pi^0p$ electroproduction off protons. 
A total
of nearly 120000 data points on unpolarized differential cross sections, longitudinally polarized beam asymmetries, and
longitudinal target and beam-target asymmetries were obtained with almost
complete coverage of the accessible phase space \cite{Az09}. The data were analyzed within the framework of two conceptually
different approaches: a) the unitary isobar model (UIM), and b) a model, employing dispersion relations
\cite{Az03,Az05}. All well established $N^*$ states in the mass range $M_{N^*}$ $<$ 1.8 GeV were 
incorporated into the $N\pi$ channel analyses.

The UIM follows the approach of ref. \cite{Dr99}. The $N\pi$
electroproduction amplitudes are described as a superposition of $N^*$ electroexcitation in s-channel and non-resonant
Born terms. A Breit-Wigner ansatz with  energy-dependent hadronic decay widths is employed 
for the resonant
amplitudes. Non-resonant amplitudes are described by a gauge invariant superposition of nucleon 
s- and u-channel exchanges, and $\pi$, $\rho$, and $\omega$
t-channel exchanges. The latter are reggeized in order to better describe the data in the second and the third resonance
regions. The final state interactions are treated as $\pi N$ rescattering in the K-matrix approximation.

In another approach, the real and imaginary parts of invariant amplitudes, that describe 
$N\pi$ electroproduction, 
are related in a model-independent way by dispersion relations \cite{Az03}. 
The analysis showed that the imaginary parts of amplitudes are dominated by resonant contributions 
at $W$ $>$ 1.3 GeV. 
In this kinematical region, they are described by resonant contributions only. 
At smaller $W$ values,
both resonant and non-resonant contributions to the imaginary part of amplitudes are taken into 
account. 

The two approaches provide good  description of the $N\pi$ data in the entire range 
covered by the CLAS
measurements: $W$ $<$ 1.7 GeV and $Q^2$ $<$ 5.0 GeV$^2$, resulting in $\chi^2$/d.p. $<$ 2.0 \cite{Az09}. 
This good description 
of a large body of different observables allowed us to obtain reliable information on $\gamma_{v}NN^*$ resonance 
electrocouplings from the analysis of $\pi^{+}n$ and $\pi^0p$ electroproduction off protons. 

The $\pi^+\pi^- p$ electroproduction data \cite{Ri03,Fe09}  provide information on 
nine independent one-fold-differential and fully-integrated cross sections in each bin of $W$ and $Q^2$ in a mass range 
$W$ $<$ 2.0 GeV, and with
photon virtualities of 0.25 $<$ $Q^2$ $<$ 1.5 GeV$^2$. Analysis of these data within framework 
of the JM reaction model \cite{Mo09,Mo07} 
allowed us to establish all essential contributing mechanisms from their manifestation in 
the measured cross sections. 
The $\pi^+\pi^- p$ electroproduction amplitudes are described in the JM model as a superposition of 
$\pi^-\Delta^{++}$, $\pi^+\Delta^{0}$, $\rho p$, $\pi^{+} D_{13}^{0}(1520)$, $\pi^{+} F_{15}^{0}(1685)$, 
$\pi^{-} P_{33}^{++}(1600)$ channels , 
and additional direct 2$\pi$ production mechanisms, 
where the
final  $\pi^+\pi^- p$ state is created without formation of unstable hadrons 
in the intermediate state. The latter mechanisms are
beyond the isobar approximation. They are required by unitarity of the $\pi^+\pi^- p$ 
amplitudes \cite{Ait1}. Direct 2$\pi$ 
production amplitudes established in the analysis the CLAS data are presented in Ref. \cite{Mo09}. 

The JM model
incorporates contributions from all well established $N^*$ states to $\pi \Delta$ and $\rho p$ isobar channels. We also
included the $3/2^{+}(1720)$ candidate state, suggested in the analysis \cite{Ri03} of the CLAS $\pi^+\pi^- p$ electroproduction data.
In the current analysis, resonant amplitudes are described using a unitarized Breit-Wigner ansatz 
proposed in Ref. \cite{Ait72}, 
and modified to make it consistent with the resonant amplitude parametrization employed in the JM model. 
This ansatz accounts for transition between the same and different  $N^*$
states in the dressed-resonant propagator, making resonant amplitudes consistent with unitarity condition.
We took into account for transitions between $D_{13}(1520)/D_{13}(1700)$, $S_{11}(1535)/S_{11}(1650)$ and $3/2^+(1720)/P_{13}(1720)$ pairs of $N^*$ states, 
and found that use of the unitarized Breit-Wigner ansatz had a minor influence on 
the $\gamma_{v}NN^*$ electrocouplings, but may affect substantially the $N^*$
hadronic decay widths.     
 Non-resonant contributions 
to $\pi \Delta$  and $\rho p$ isobar channels are described in \cite{Mo09} and \cite{Mo071}, respectively.  
Other isobar channels are described by non-resonant amplitudes presented in Refs.~\cite{Mo07,Mo05}.

The JM model provided reasonable description of $\pi^+\pi^- p$ differential cross sections at $W$ $<$ 1.8 GeV and $Q^2$ $<$ 1.5 GeV$^2$ 
with $\chi^2$/d.p. $<$ 3.0.The successful description of  $\pi^+\pi^-p$ electroproduction  cross sections
allows us to isolate the resonant parts and to determine  $\gamma_{v}NN^*$  
electrocouplings, as well as the $\pi \Delta$ and $\rho p$ decay widths. 

\section{Results on the $\gamma_{v}NN^*$ electrocouplings and impact on the studies of $N^*$ structure}

\begin{figure}
  \includegraphics[height=.24\textheight]{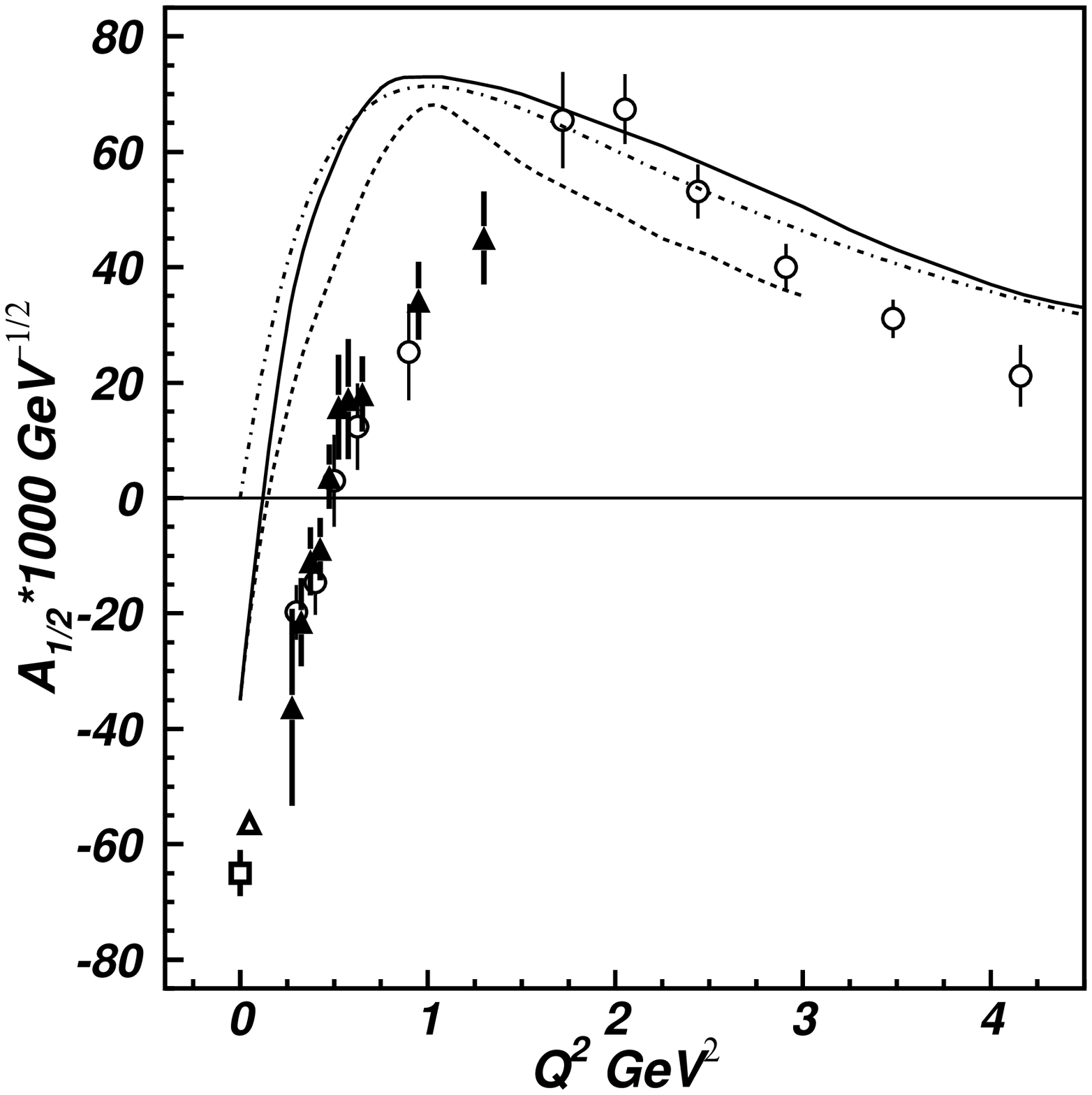}
  \includegraphics[height=.24\textheight]{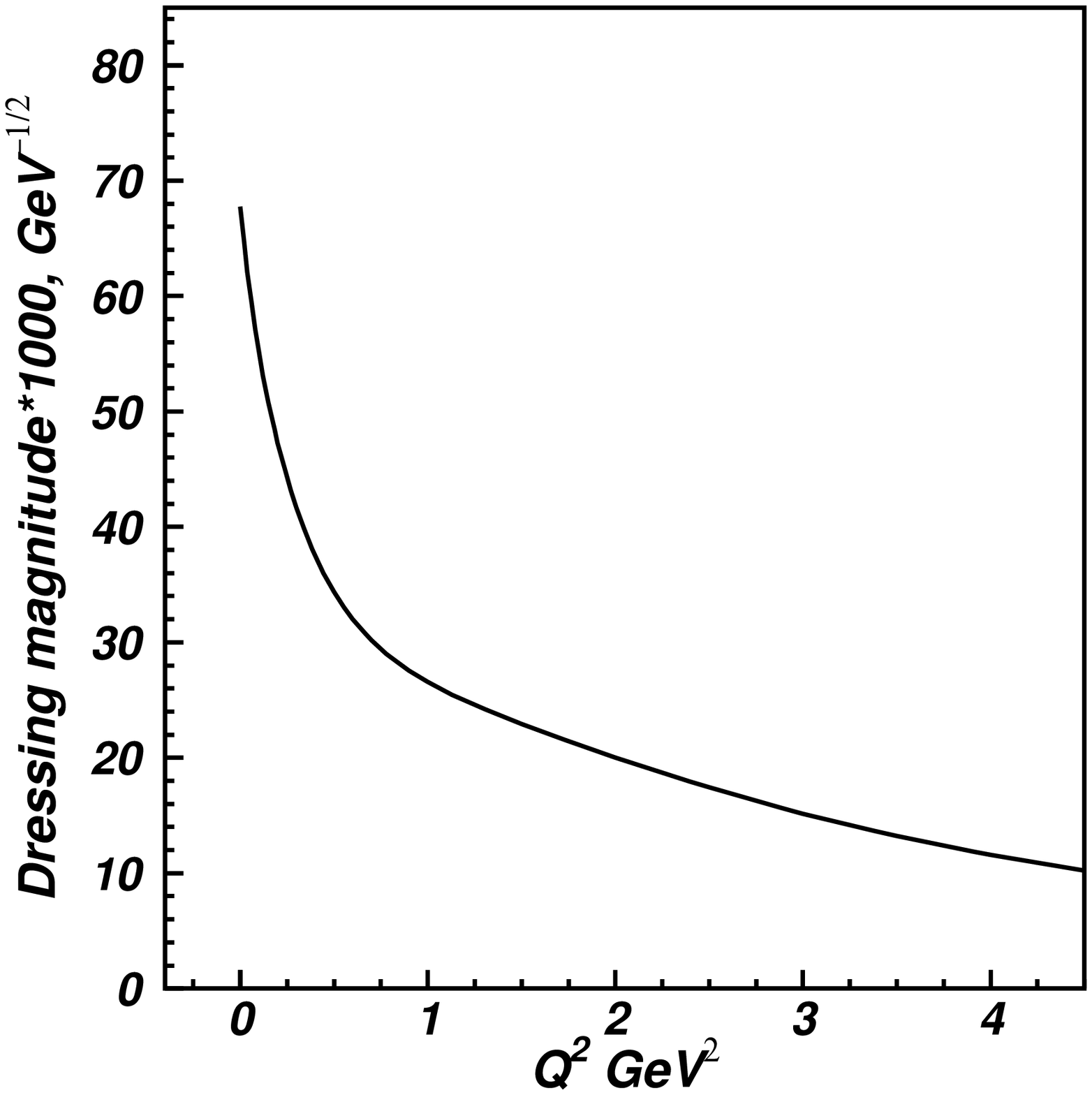}
  \caption{Left: Electrocouplings of the $P_{11}(1440)$ resonance determined in independent analyses of the CLAS data on $N\pi$ (circles) and $\pi^+\pi^- p$ (triangles)
  electroproduction off protons. Square and triangle at $Q^2$=0 correspond to RPP \cite{rpp} and the CLAS $N\pi$  \cite{Dug09} photoproduction
  results, respectively. The results of relativistic light-front quark models \cite{Az07,Ca95} are shown by solid and dashed
  lines, respectively. Results of the covariant valence quark-spectator diquark model \cite{Ra10} are shown by the dashed dotted line. 
  Right: Estimate for absolute values of meson baryon dressing amplitude contributing to $A_{1/2}$ electrocoupling obtained within
  the framework of coupled-channel model \cite{Lee08} from a global fit of the data on $N\pi$  photo-, electro-, and
  hadroproduction.}
  \label{p11}
\end{figure}

\begin{figure}
  \includegraphics[height=.24\textheight]{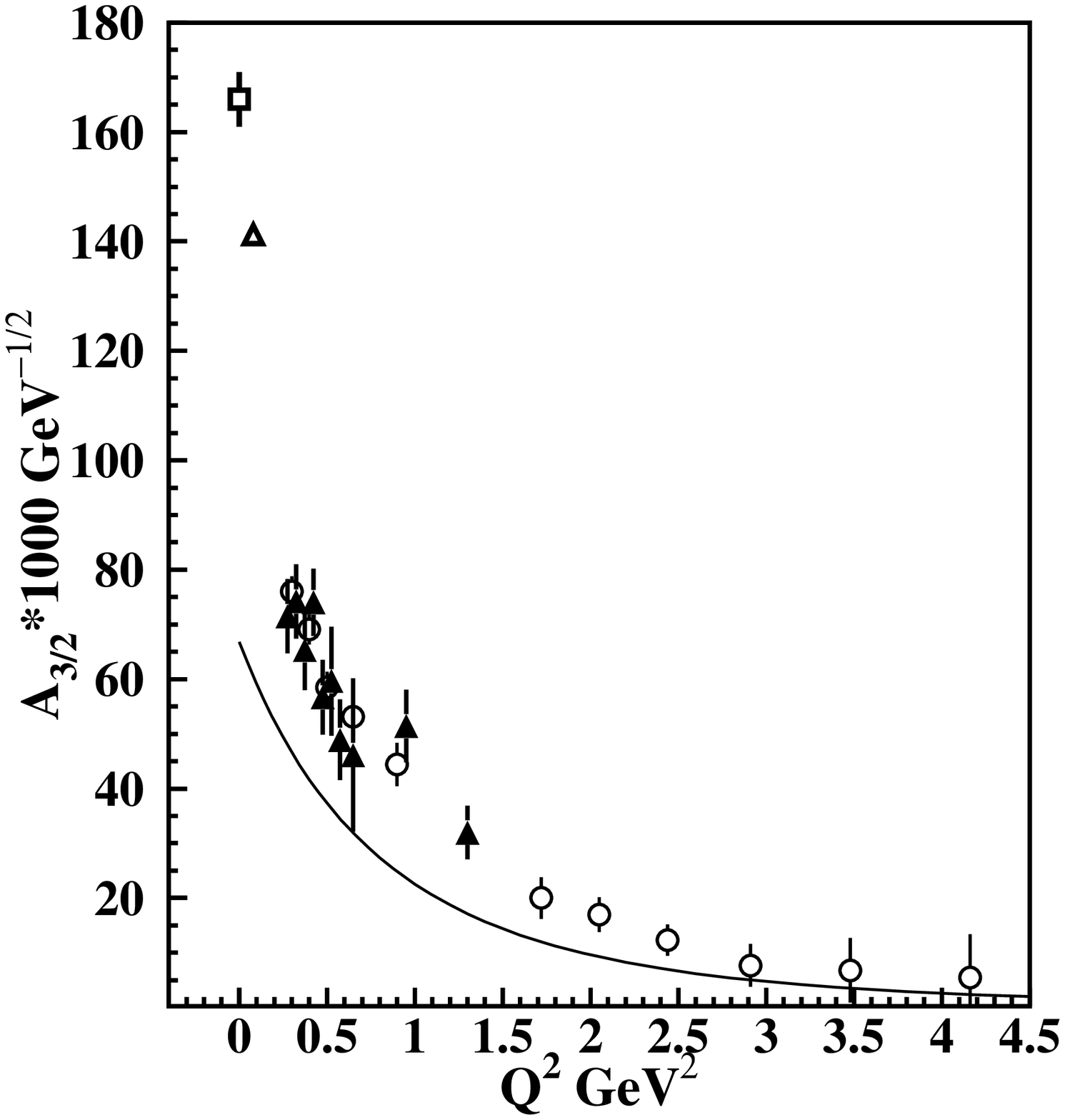}
  \includegraphics[height=.24\textheight]{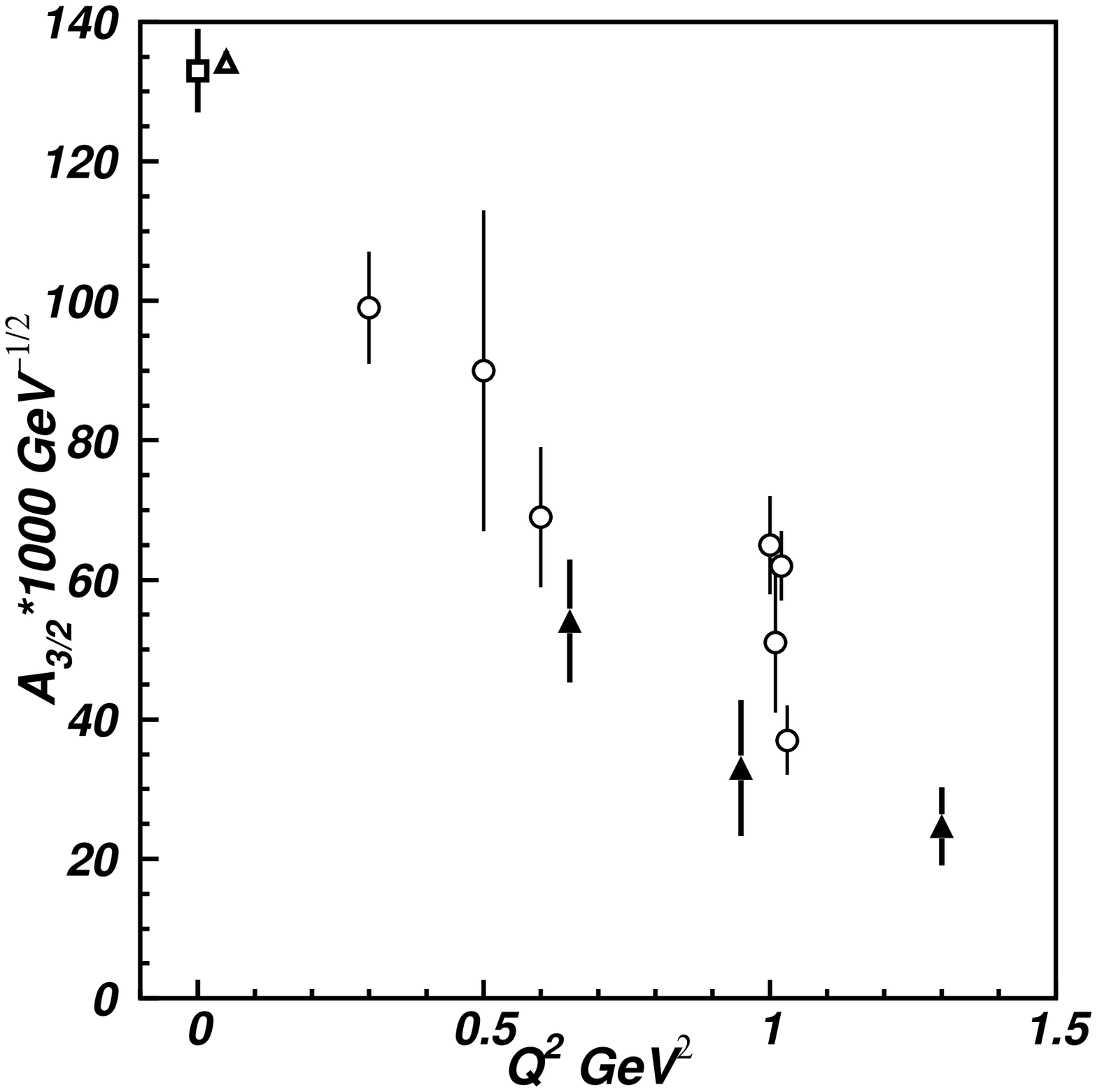}
  \caption{Electrocouplings of $D_{13}(1520)$ (left) and $F_{15}(1685)$ (right) resonances determined in independent 
  analyses of the CLAS \cite{Az09} and world \cite{Bu03} data on $N\pi$, and the CLAS data on $\pi^+\pi^- p$  \cite{Ri03,Fe09}
  electroproduction off protons. Symbols are the same as in Fig.~\ref{p11}. 
   The results of quark model \cite{Gia} are shown by solid line.}
  \label{d13f15}
\end{figure}

Electrocouplings of $P_{11}(1440)$, $D_{13}(1520)$ and $F_{15}(1885)$ excited proton states 
are shown in Figs.~\ref{p11} and~\ref{d13f15}. The results from $N\pi$ and $\pi^+\pi^-p$ channels are consistent within
their uncertainties. Consistent results on $\gamma_{v}NN^*$ electrocouplings 
for several excited proton states determined in
independent analyses of major meson electroproduction channels with different backgrounds
demonstrate that the reaction models described above provide reliable 
evaluation of these fundamental quantities. It makes possible to determine 
electrocouplings  of all resonances that decay preferentially 
to the either $N\pi$ or $N\pi\pi$ final states analyzing independently the $N\pi$ or 
$\pi^+\pi^-p$ electroproduction channels, respectively. The studies of $N\pi$ exclusive channels 
are the primary source of information on electrocouplings of the $N^*$ states with masses below 
1.6 GeV \cite{Az09}.  The $\pi^+\pi^-p$ electroproduction off protons allowed us to check
the results from $N\pi$ data analyses for the resonances that have substantial decays to both $N\pi$
and $N\pi\pi$ channels. Furthermore, they are of particular 
importance for the evaluation of  high-lying resonance electrocouplings, 
 since most $N^*$'s with masses above 1.6 GeV  decay
 preferentially via two pion emission.

\begin{figure}
  \includegraphics[height=.2\textheight]{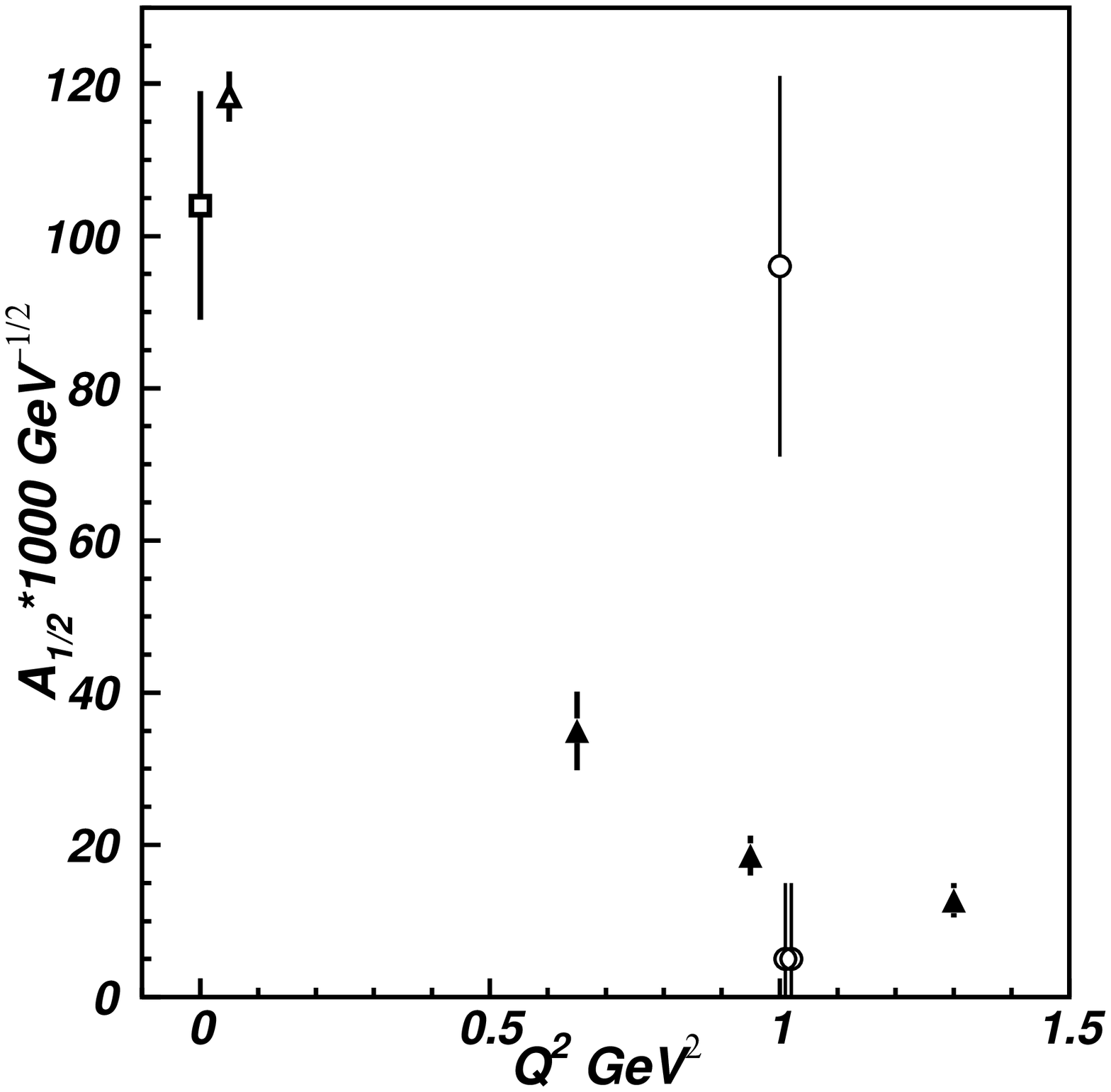}
  \includegraphics[height=.2\textheight]{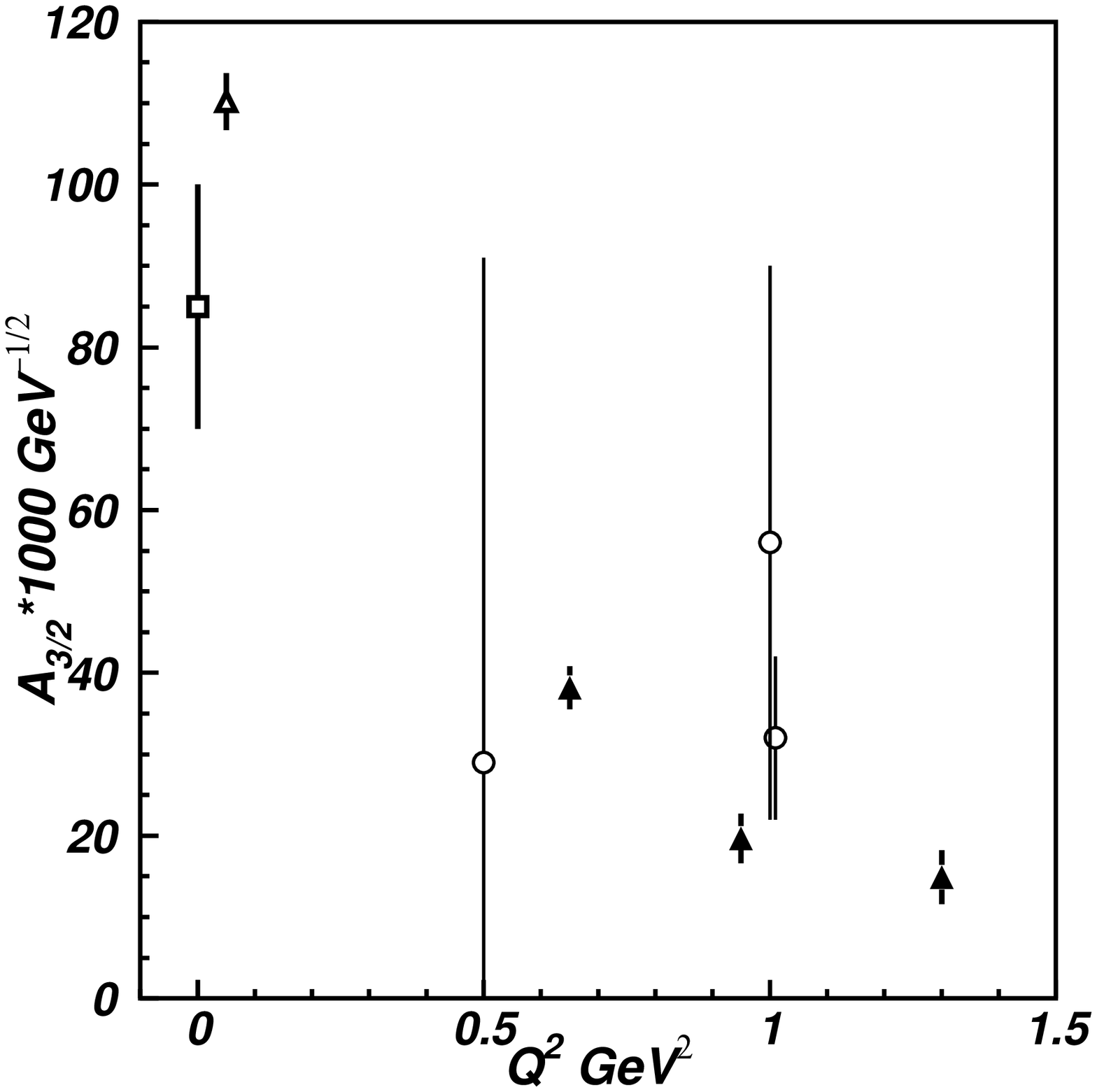}
  \includegraphics[height=.2\textheight]{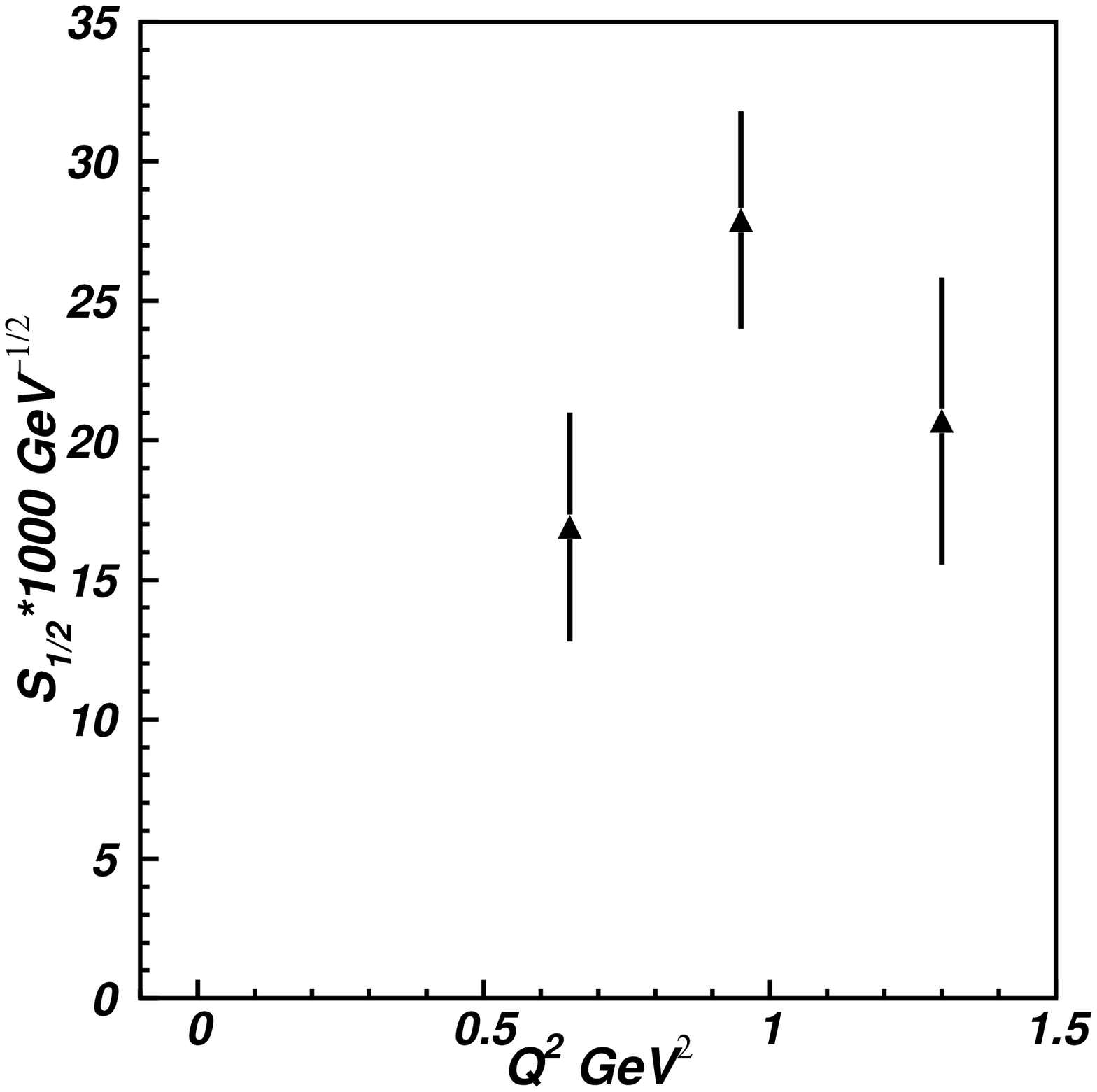}
  \caption{Electrocouplings of $D_{33}(1700)$ resonance $A_{1/2}$ (left), $A_{3/2}$ (middle) and $S_{1/2}$ (right)  
  determined in independent 
  analyses the CLAS data on $\pi^+\pi^- p$  \cite{Ri03,Fe09} and world data \cite{Bu03} on $N\pi$
  electroproduction off protons. Symbols are the same as in Fig.~\ref{p11}.}
  \label{d33}
\end{figure}

 Preliminary results on electrocouplings of $S_{31}(1620)$, $S_{11}(1650)$, $F_{15}(1685)$, $D_{33}(1700)$ and $P_{13}(1720)$ 
 states were
 obtained from the CLAS $\pi^+\pi^-p$ electroproduction data \cite{Ri03}. Electrocouplings 
 of the $D_{33}(1700)$
 state determined from  $N\pi$ (previously available world data \cite{Bu03}) and $\pi^+\pi^-p$ (the CLAS results) electroproduction channels 
 are shown in Fig.~\ref{d33}. 
 The CLAS results improved considerably our knowledge of the $D_{33}(1700)$ 
 electrocouplings. They provided accurate data on the $Q^2$-evolution of the transverse 
 electrocouplings and the first information on the longitudinal
 electrocouplings of all the above mentioned excited proton states.

.

Analysis of the CLAS results on $P_{11}(1440)$ electrocouplings revealed major features 
of this state structure 
that remained a mystery for decades.
Two light-front quark models \cite{Az07,Ca95} and conceptually different covariant valence quark-spectator diquark model
\cite{Ra10} provided reasonable descriptions of  $P_{11}(1440)$ electrocouplings at $Q^2$ $>$ 1.5 GeV$^2$. 
In these models, the $P_{11}(1440)$
structure is described as the first radial excitation of
three-quark (3$q$) ground state. The CLAS results showed that at $Q^2$ $>$ 1.5 GeV$^2$ the $P_{11}(1440)$ structure is
determined mostly by a core of three constituent quarks. However, at $Q^2$ $<$ 1.0 GeV$^2$, 
quark models fail to
describe the $A_{1/2}$ electrocoupling. This is an indication of additional contributions beyond those from a quark core.
 A general unitarity condition requires  the contribution to $\gamma_{v}NN^*$ electrocouplings 
 from meson-baryon dressing when the $N^*$
 state is excited through a sequence of the following processes: a) non-resonant meson-baryon production by a virtual photon 
 and b) resonance formation in subsequent meson-baryon
 scattering \cite{Lee10}.  The absolute value of meson-baryon dressing amplitudes determined from the data
on $N\pi$ photo-, electro- and hadroproduction \cite{Lee08} is shown in Fig~\ref{p11}. 
It is maximal at small photon
virtualities and may, in part, be responsible for the differences between quark model expectations and the CLAS results on
 $P_{11}(1440)$ electrocouplings.

 Electrocouplings of the $D_{13}(1520)$ state, shown in Fig.~\ref{d13f15}, are well described at $Q^2$ $>$ 2.0 GeV$^2$ 
 within the framework of quark model \cite{Gia}, which assumes the contributions from three constituent quarks in the first orbital
 excitation with $L$=1. This model underestimates the $A_{3/2}$ electrocoupling at $Q^2$ $<$ 2.0 GeV$^2$. 
 At these photon virtualities absolute values of meson-baryon dressing contributions to this electrocoupling are maximal
 \cite{Lee08}. Differences between the CLAS
 results on $D_{13}(1520)$ electrocouplings and expectations of quark model \cite{Gia} may be 
 related to the meson-baryon cloud.
 
 We conclude that the structure of excited proton states with masses below 1.6 GeV determined by combined
 contributions from the internal core of three constituent quarks and the external meson-baryon cloud.

 \section{Conclusion }

 The information on $Q^2$ evolution of $\gamma_{v}NN^*$ electrocouplings of many excited proton states in 
 the mass range up to 1.8 GeV 
 has become available from the analyses of exclusive $N\pi$ and $\pi^+\pi^-p$ electroproduction off protons measured with 
 CLAS detector. Consistent results obtained from independent analyses of major meson
 electroproduction channels demonstrate reliable extraction of these fundamental quantities. 
 They open up new 
 opportunities for studies of the non-perturbative strong interaction that is responsible for the 
 formation of excited proton states of different quantum
 numbers. In particular, they stimulated the development of $N^*$ 
 structure models presented, in part, at this workshop \cite{Ra10,Az11,San11,Ra11,GdT11}. Furthermore, 
 two conceptually different  approaches of QCD-Lattice QCD \cite{Li09,Lin11,Ed11,Ed11p,Br11,Br09} and
 Dyson-Schwinger equations \cite{CRo11,ICl11}-are making progress toward the description of 
 $\gamma_{v}NN^*$ electrocouplings 
 from the first  principles of QCD.


\begin{theacknowledgments}
This work was supported
in part by the U.S. Department of Energy, the Russian Federation
Government Grant 02.740.11.0242, 07.07.2009, and the Department of Education and Science 
of Republic of Armenia Grant-11-1C015, the Skobeltsyn Institute of Nuclear Physics and
Physics Department at Moscow State University, Yerevan Physics Institute (Armenia). 
Jefferson Science Associates, LLC, operates Jefferson Lab
under U.S. DOE contract DE-AC05-060R23177. 
\end{theacknowledgments}



\bibliographystyle{aipproc}   

\bibliography{sample}

\begin{thebibliography}{9}

\bibitem{Bu11} V.~D. Burkert et al.,  
J. Phys: Conf. Ser. $\bf{299}$, 012008
(2011).

\bibitem{CL12} I.~G.~Aznauryan, et al., arXiv:0907.1901 [nucl-th].




\bibitem{CRo11} C.~D.~Roberts, this proceedings.
\bibitem{Az09} I. G. Aznauryan et al., CLAS Collaboration,
Phys. Rev. C $\bf{80}$, 055203 (2009).
\bibitem{Az03} I. G. Aznauryan, Phys. Rev. C $\bf{67}$, 0152009 (2003).
\bibitem{Az05} I. G. Aznauryan et al.,
Phys. Rev. C $\bf{71}$, 015201 (2005).
\bibitem{Dr99} D. Drechsel et al.,
Nucl. Phys. A $\bf{645}$, 145 (1999).
\bibitem{Ri03} M.~Ripani et al., CLAS Collaboration, Phys. Rev. Lett. {\bf
91}, 022002 (2003)  



\bibitem{Fe09} G.~V. Fedotov et al., CLAS Collaboration, 
Phys. Rev. C{\bf 79}, 015204 (2009).



\bibitem{Mo09} V.~I.~Mokeev et al., Phys. Rev.C {\bf 80}, 045212 (2009).

\bibitem{Mo07} V. I. Mokeev et al., in "Proceedings of the 11th Workshop on
the Physics of Excited Nucleons. NSTAR2007", Springer 2008, ed. by H-W. Hammer, V.Kleber, U.Thoma, H.
Schmieden, p. 76. 
\bibitem{Ait1} I.~J.~R~Aitchison and J.~J.~Brehm,  Phys. Rev. D {\bf 17}, 3072 (1978). 
\bibitem{Mo071} V.~I.~Mokeev et al., Phys. of Atom. Nucl. {\bf 70}, 427 (2007).




\bibitem{Ait72} I.~J.~R.~Aitchison,  Nucl. Phys. A {\bf 189}, 417 (1972). 




\bibitem{Mo05} V. I. Mokeev et al., in "Proc. of the Workshop on
the Physics of Excited Nucleon. NSTAR2005", ed. by S.Capstick, V.Crede,
P.Eugenio, World Scientific Publishing Co.,hep-ph$/$0512164, p. 47.



\bibitem{rpp} Review of Particle Physics, J. Phys. G {\bf 37}, 075021 (2010).
\bibitem{Dug09} M. Dugger et al., CLAS Collaboration,  Phys. Rev. C {\bf 79}, 065206 (2009).

\bibitem{Az07} I. G. Aznauryan,
Phys. Rev. C $\bf{76}$, 025212 (2007).
\bibitem{Ca95} S. Capstick and B. D.
Keister, Phys. Rev. D $\bf{51}$, 3598 (1995).
\bibitem{Ra10} G.~Ramalho and K. Tsushima,  Phys. Rev. D {\bf 81}, 074020 (2010). 

\bibitem{Lee10} N. Suzuki, T. Sato, T.-S.H. Lee, Phys. Rev. C $\bf{82}$, 045206 (2010), 
arXiv:1006.2196 [nucl-th].
\bibitem{Lee08} B. Julia-Diaz, et al., Phys. Rev. C $\bf{77}$, 045205 (2008).
\bibitem{Bu03} V.~D.~Burkert et al.,  Phys. Rev. C {\bf 67}, 035204 (2003).
\bibitem{Gia} M. Aiello, M. M. Giannini and E. Santopinto, J.  Phys. G {\bf 24}, 753 (1998).


\bibitem{Az11}  I.~G.~Aznauryan  and V.~D.~Burkert, this proceedings
\bibitem{San11} E..~Santopinto, this proceedings.
\bibitem{Ra11}  G..~Ramalho, this proceedings.
\bibitem{GdT11}  Guy de Teramond, this proceedings.

\bibitem{Li09} H.-W. Lin, Chin. Phys. C $\bf{33}$, 1238 (2009).
\bibitem{Lin11} H.-W.~Lin, this proceedings.
\bibitem{Ed11}  R.~G.~Edwards, this proceedings.
\bibitem{Ed11p}  R.~G.~Edwards, et al., for the Hadron Spectroscopy Collaboration, arXiv:1104.5152 [hep-ph].
\bibitem{Br11}  V.~Braun, this proceedings.
\bibitem{Br09} V.~Braun, et al., Phys. Rev. Lett. $\bf{103}$, 072001 (2009).


\bibitem{ICl11}  I.~Cloet, this proceedings.

\end{thebibliography}

\IfFileExists{\jobname.bbl}{}
 {\typeout{}
  \typeout{******************************************}
  \typeout{** Please run "bibtex \jobname" to optain}
  \typeout{** the bibliography and then re-run LaTeX}
  \typeout{** twice to fix the references!}
  \typeout{******************************************}
  \typeout{}
 }


\end{document}